\documentclass[journal]{IEEEtran}
\usepackage {graphicx}
\usepackage{subfigure}
\DeclareGraphicsExtensions{.eps,.pdf,.jpeg,.png}
\usepackage[cmex10]{amsmath}
\usepackage{amssymb}
\usepackage{cite}
\usepackage{multirow}
\usepackage{array} 
\begin{document}

\title{A multi-level soft frequency reuse technique for wireless communication  systems}

\author{Xuezhi Yang,~\IEEEmembership{Senior Member,~IEEE }

\thanks{Xuezhi Yang is the inventor of soft frequency reuse and several other key technologies of 3G and 4G.  He is now a freelance researcher on wireless communications, Beijing, China (email: yangxuezhi.ieee@gmail.com) }% <-this % stops a space
\thanks{}}

% The paper headers
\markboth{}{}

% make the title area
\maketitle

%\IEEEpeerreviewmaketitle

\begin{abstract}
A multi-level soft frequency reuse (ML-SFR) scheme and a resource allocation methodology are proposed for wireless communication systems in this letter.  In the proposed ML-SFR scheme, there are 2N power density limit levels, achieving better interference pattern and further improving the cell edge and overall data rate, compared to the traditional 2-level SFR scheme. The detailed design of an 8-level SFR scheme is demonstrated. Numeral results show that the cell edge spectrum efficiency is increased to 5 times of that of reuse 1 and  the overall spectrum efficiency  is improved by 31\%. ML-SFR can be utilized in the current 4G system and would be a candidate key technology for future 5G systems. 
\end{abstract}

\begin{IEEEkeywords}
Multi-level soft frequency reuse, inter cell interference coordination,  interference pattern, LTE, 5G.
\end{IEEEkeywords}

%\IEEEpeerreviewmaketitle

\section{Introduction}

\IEEEPARstart{S}{oft} frequency reuse (SFR)  \cite{YangSFRPatent2006, HuaweiSfr2005} has become one of most important enablers for wireless systems, especially orthogonal frequency division multiplexing (OFDM) based ones, to achieve  high data-rate communications. It has been extensively  explored under the subject of \textit{inter cell interference coordination} (ICIC) in 3GPP LTE \cite{RongZhang2010}. It is also adopted by the Wimax standard, however given another name\textit{ fractional frequency reuse} \cite{MobileWiMAXWhitePaper1}, a term originally meaning a specific configuration of directional antennas \cite{FaruqueFFR1998}, and later used by \cite{ KhandekarUSPatentFFR2005} to represent \textit{reuse partitioning} \cite{HalpernReusepartitioning,Sternad2003}.

\begin{figure}[!ht]
\begin{center}
\includegraphics[bb=0.5in  0.5in 8in 4in, width=2.8in]{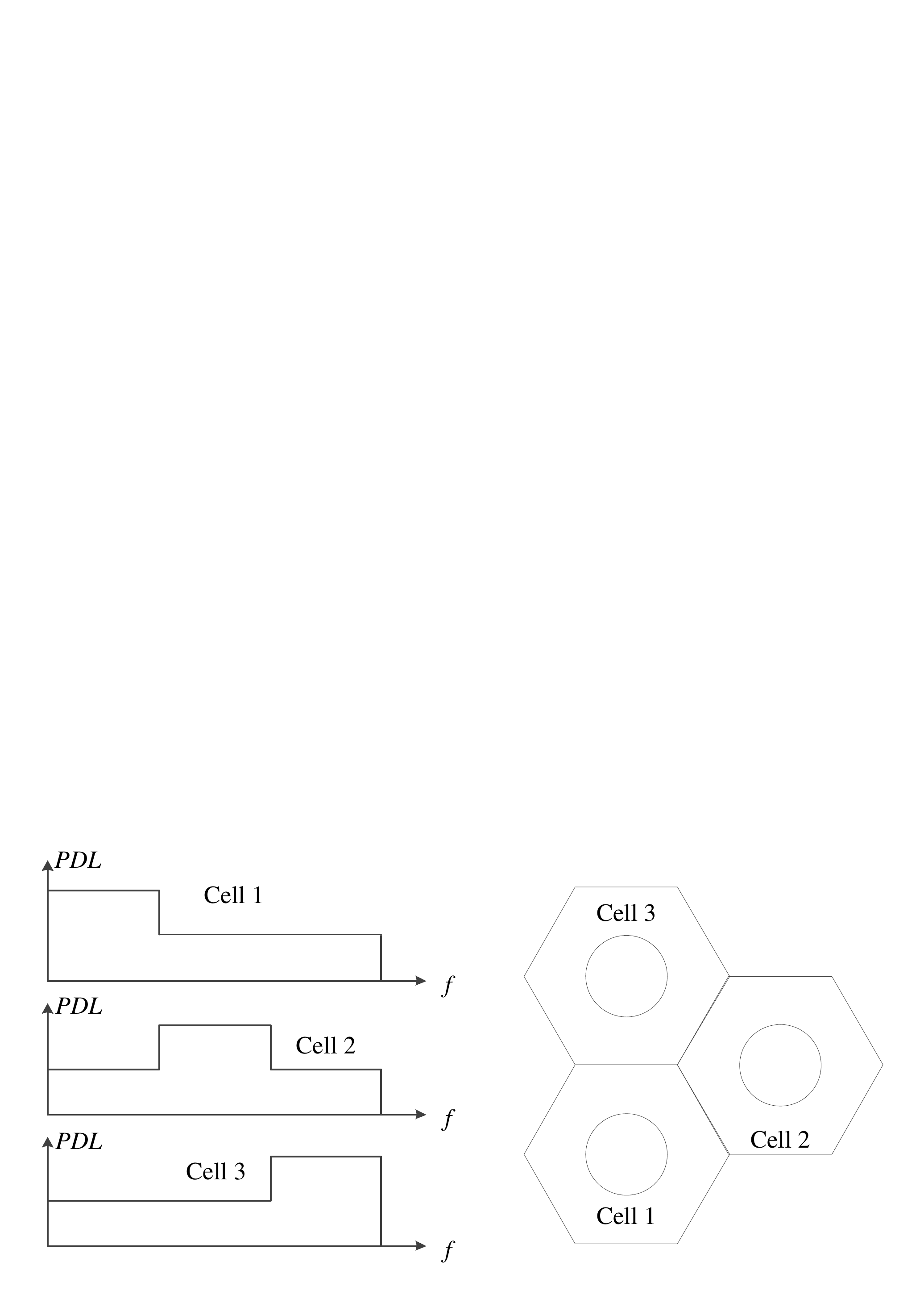}
\caption{Power density upper limit and coverage of soft frequency reuse.}
\label{Fig-SFR}
\end{center}
\end{figure}

In an example SFR configuration, as illustrated in Fig. \ref{Fig-SFR}, the whole bandwidth is divided into three parts, in which one part, called primary band, has  a higher power density upper limit (PDL) than the other two, called secondary bands.  The primary bands of adjacent cells are orthogonal to and do not interfere with each other. 

SFR is closely related to resource allocation, or specifically scheduling in LTE system \cite{MaoXuehongAdaptiveSFR2008, BohgeSFR2009, Doppler2009}. In fact, SFR defines a network-level framework for resource allocation in each cell.  Due to the constrains of SFR, a cell edge UE can only be allocated resources in the primary band while a cell centre UE can access the whole bandwidth. Because the primary bands of adjacent cells are orthogonal to each other, severe interference  is avoided at cell edge, improving the cell edge data rate compared to the reuse 1 system \cite{YangSFRPatent2006, HuaweiSfr2005}.   SFR  can be straightforwardly extended to the time domain  to make a \textit{soft time reuse} scheme \cite{YangSTRPatent2005}, or enhanced inter cell interference coordination (eICIC) in 3GPP \cite{eICIC3GPP}.

An important  parameter for SFR is the ratio $\gamma$ expressed by

\begin{equation}
\label{Eq-ratio}
\gamma=\frac{\textrm{PDL of secondary  band}}{\textrm{PDL of primary  band}}.
\end{equation}
It has been pointed out in \cite{HuaweiSfr2005} that when  $\gamma$ increases, the cell edge capacity decreases and the cell centre capacity increases, and vise versa. It is clear that when all the traffic happens at the cell edge/centre,  $\gamma$ should be 0/1. However, when the traffic is uniformly distributed, the optimal value for   $\gamma$ is still an open problem. Another question is, since the widely accepted two-PDL-level SFR (SFR-2) scheme can improve the cell edge data rate, could  more PDL levels help to further improve the performances? Even there are some tentative moves in \cite{BohgeSFR2009} and \cite{Doppler2009}, no substantial progress has been made on this topic.

In this letter, a multi-level soft frequency reuse (ML-SFR) technique is proposed, in which the whole frequency band is divided into several parts. On each part,  a SFR-2 scheme with its specific $\gamma$ value is employed, based on the discovery that the optimal $\gamma$ value is closely related to UE position.  The  closer a UE is to the cell edge, the  larger transmit power it  requires,  and the lower interfering power from adjacent cells it can endure, corresponding to a larger  $\gamma$ value.  A resource allocation methodology based on ML-SFR is proposed. ML-SFR and the proposed resource allocation methodology  further optimize the interference pattern and  improve the cell-edge and overall  data rate, compared to SFR-2. 

The rest of this letter is organized as follows.  Section II gives the problem formulation.  A ML-SFR scheme and  a resource allocation methodology are proposed in Section III and Section IV, respectively. Section V demonstrates the detailed design of an 8-level SFR (SFR-8) scheme and numerical results. Section VI concludes this letter.

\section{Problem Formulation}

\begin{figure}[!ht]
\begin{center}
\includegraphics[bb=0.5in  0.5in 8in 4in,width=2.2in]{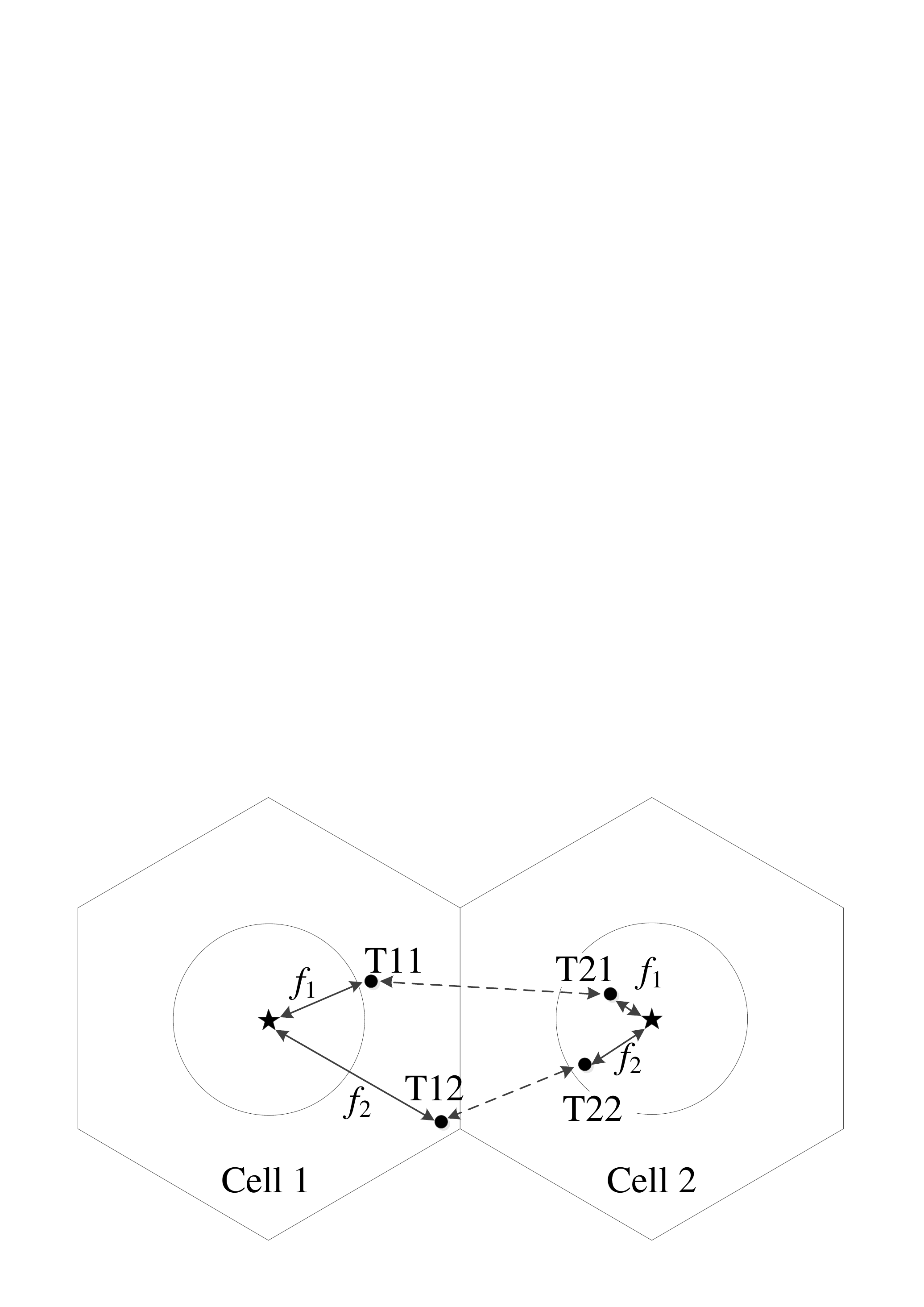}
\caption{An interference pattern of the SFR-2  scheme.}
\label{Fig-eSFR018}
\end{center}
\end{figure}

Let us have a look at an interference pattern of the SFR-2 scheme, illustrated in Fig. \ref{Fig-eSFR018}. There are two UEs in the cell edge area of Cell 1, denoted as T11 and T12, communicating with their base station using frequency $f_1$ and $f_2$ in the primary band, either in downlink or uplink.  There are two UEs in the cell centre area of Cell 2, denoted as T21 and T22, communicating with their base station also on frequency $f_1$ and $f_2$. According to the SFR definition,  $f_1$ and $f_2$ should be in the secondary band of Cell 2.  So, the interference pattern is, T11 interferes with T21 and T12 interferes with T22, as shown in  Fig. \ref{Fig-eSFR018}.

In this example, T12 and T22 are more far away from their base station than T11 and T21, respectively, so requiring bigger  transmit power.  In such a case, switching the communication resources for T21 and T22 will make a better interference pattern. Since T12 is  more vulnerable  than T11, it is appropriate to pair it with T21, which has smaller interfering power than T22. With such a pattern,  the UE at the most cell edge can achieve higher data rate at the cost of lower data rate of the UE at the most cell centre. This is desirable for operators since user complains can be effectively reduced by increasing the data rate of the most vulnerable UEs.

However, SFR-2 can not provide enough constraints to always realize the better interference pattern. In a 2-cell scenario as in Fig. \ref{Fig-eSFR018},  the better pattern can be achieved by the chance of 50\% via random resource allocation. In a 7-cell scenario, one central cell plus 6 surrounding cells, the chance of random allocation to realize the optimal pattern falls to $(1/2)^6=1/32$. In order to  optimize the interference pattern, a multi-level soft frequency reuse is presented, as follows.

\section{Multi-level soft frequency reuse}

\begin{figure}[!ht]
\begin{center}
\includegraphics[bb=0.5in  0.5in 8in 4in, width=2.5in]{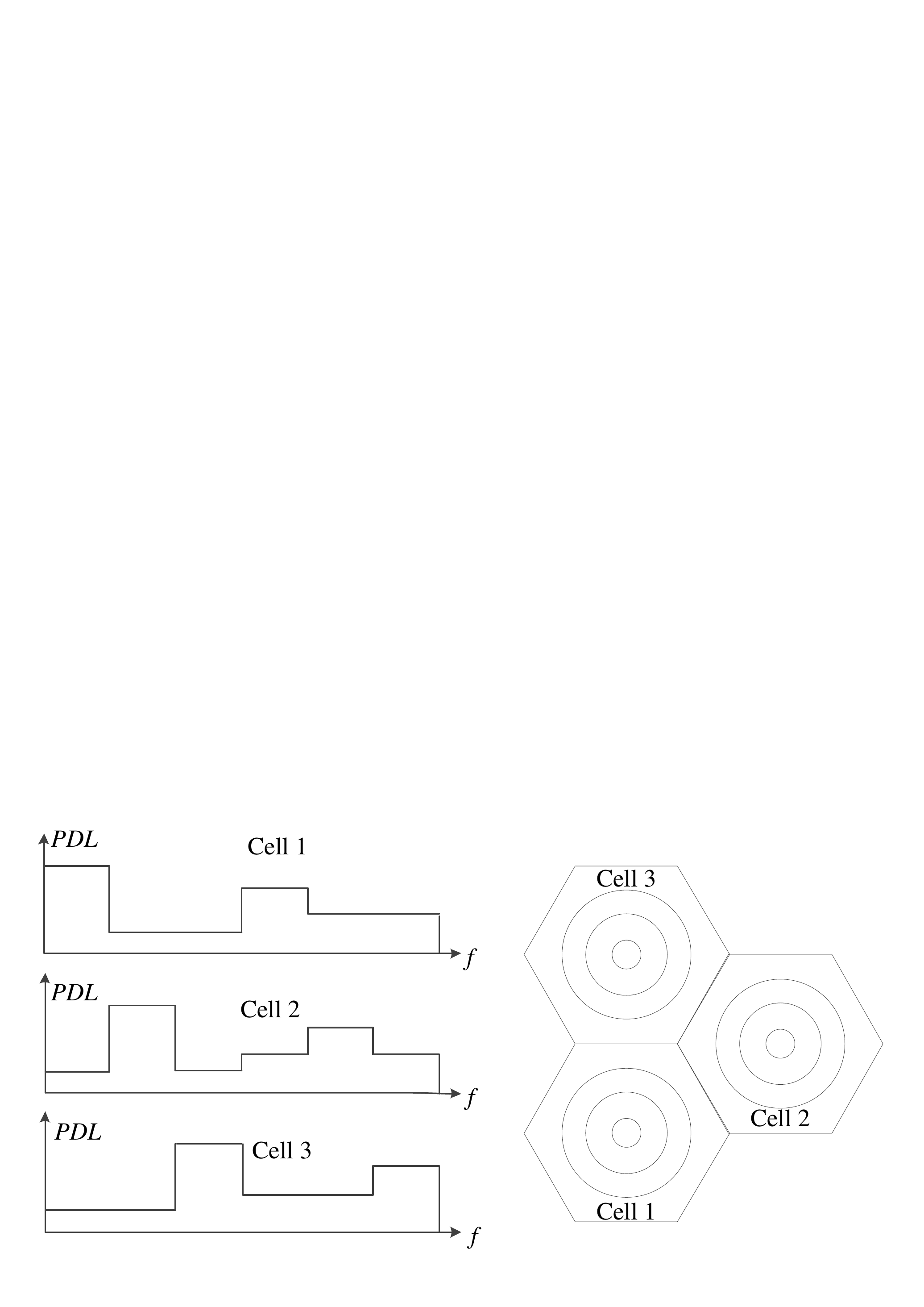}
\caption{Power density upper limit and coverage of a SFR-4 scheme.}
\label{Fig-ML-SFR}
\end{center}
\end{figure}

In a ML-SFR scheme, the whole frequency band is divided into $N$ parts on each of which a separate SFR-2 scheme is employed.  In Cell $i$, the PDLs of the primary and secondary bands of part $n\in [1,N]$ are denoted as $h_n^{(i)}$ and $l_n^{(i)}$ respectively, satisfying

\begin{equation}
\label{ }
l_1^{(i)}\leqslant l_2^{(i)}\leqslant \cdots \leqslant l_N^{(i)}\leqslant h_N^{(i)}\leqslant \cdots\leqslant h_2^{(i)}\leqslant h_1^{(i)}.
\end{equation}

A  SFR-4 scheme is shown in Fig.  \ref{Fig-ML-SFR}. It is worth noticing the  specific PDL setting  in which the highest PDL pairs with the lowest, the second highest pairs with the second lowest, and so on. 

In a SFR-2 scheme, the whole cell is quantized into two areas, cell centre and cell edge. This is a relatively coarse framework for resource allocation.   In a ML-SFR scheme, there are totally $2N$ PDL levels, dividing the whole cell into $2N$ areas and making a more refined framework. 

\section{Resource allocation based on multi-level soft frequency reuse}

Resource allocation happens at the instance of call access in voice communication systems like GSM. In LTE system, it happens periodically in the process of data transmission, called scheduling. Similar to SFR-2, ML-SFR also defines a network-level framework for resource allocation in each cell.

It is obvious  a band with higher PDL have a larger coverage. Generally, a UE can be assigned the resources in the frequency band that covers it. However,  we suggest the methodology of allocating resources  to a UE in the bands with the possible smallest coverage to optimize the system performances.

First of all, a coverage area is determined for each of the frequency bands according to their PDLs. The relationship between the PDL and coverage is  an implementation problem  which  operators would optimize based on the realistic scenarios. Each UE reports its position to the base station and a list including the bands covering the UE is created. When allocating resources to a UE, the allocator search  the band with the smallest coverage in the list for available resources. If there are not enough available resources, then search the band with the second smallest coverage, and so on, till enough resources are found or a negative conclusion is drawn.

Let us  revisit the example in Fig. \ref{Fig-eSFR018} assuming a SFR-4 scheme is used.  Suppose $f_1$ and $f_2$ belong to the primary bands  of Cell 1 and $f_1$ has a smaller PDL than $f_2$, then according to the definition of ML-SFR, they  belong to the secondary bands of Cell 2 and  $f_1$  has a greater PDL than $f_2$.  If  there is enough separation in distance between the UEs  in Fig. \ref{Fig-eSFR018}, a list including the bands with the coverage in ascending order  could be established for each UE as in Tab. \ref{BandList}. Then the  resource assigned to each UE is the first frequency in  the list, achieving the better interference pattern aforementioned.

\begin{table}[ht]
  \centering 
  \caption{ }\label{BandList}
\begin{tabular}{c|c|c|c|c}
  \hline
   UE&T11& T12& T21& T22\\  \hline
   List&$f_1, f_2$ &$f_2$&$f_2, f_1$&$f_1$\\
\hline 
\end{tabular}
\end{table}

From this example we can also learn the rationality of the suggested resource allocation methodology. If T11 is assigned $f_2$ with the larger coverage than $f_1$, there will no resources left for T12 while $f_1$ is still spare.

It is worth noticing  that ML-SFR and the suggested resource allocation methodology can be employed in any cellular systems without the need of standardization works. From the above discussion we can learn that  UE's position is the only measurement needed by the base station to  implement ML-SFR, which is already realized by all cellular systems to support handover function.  Commonly,   a UE measures the received power of a beacon signal send by the base station and reports it back as  the indication of its distance to the base station.  In LTE, the  common reference signal (CRS) plays the role of the beacon signal. 

In LTE, a common way to employ SFR-2  on one carrier is  to choose a part of subcarriers as primary band and others as secondary bands. ML-SFR can be implemented in the similar way by dividing a carrier into more parts. However, since CRS and physical downlink control channel (PDCCH) are distributed throughout the whole carrier and are transmitted with full power \cite{Stefania}, they will interfere the primary bands of adjacent cells and impair the gain of ML-SFR. 

In order to overcome the CRS and PDCCH interference problem, an alternative way is to use multiple carriers, each carrier as a band with its specific PDL. Then the power of  CRS and PDCCH on each carrier can be reduced along with the traffic channels.   A drawback for an operator with limited spectrum is the limited peak data rate, because  adopting the multi-carrier solution means one wideband carrier is divided into several narrowband carriers and a LTE UE can only transmit data on one carrier.   In LTE-A, with the deployment of carrier aggregation \cite{eICIC3GPP} , one UE can transmit data on several component carriers, the peak data rate problem can be solved.

\section{Design of Multi-level soft frequency reuse}

\begin{figure}[!ht]
\begin{center}
\includegraphics[bb=0.5in  0.5in 6.5in 6.5in, width=1.5in]{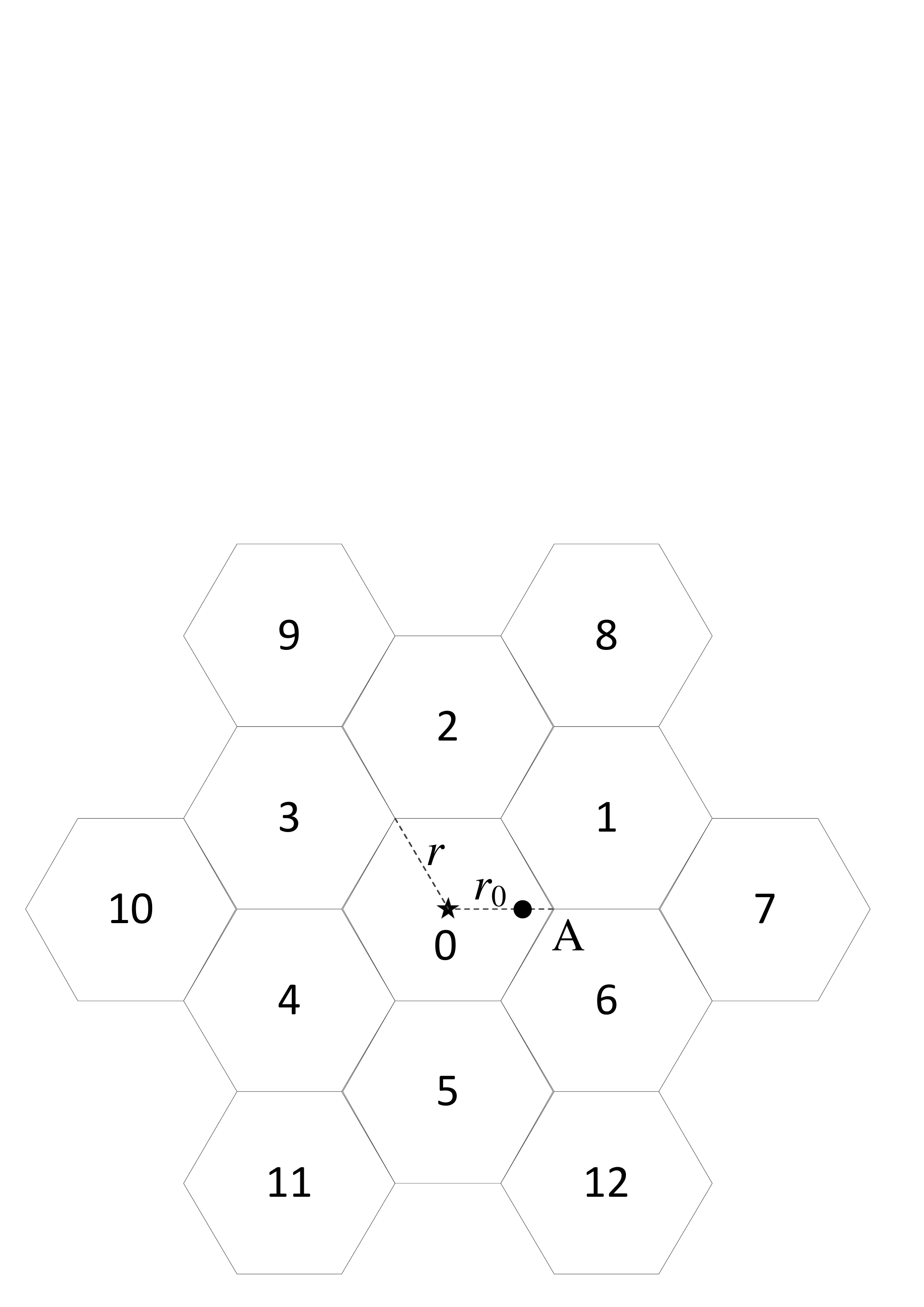}
\caption{A cellular network including 13 cells.}
\label{Fig-SFRTopo}
\end{center}
\end{figure}

Consider a cellular network including 13 cells with  radius  $r$,  as shown in Fig. \ref{Fig-SFRTopo}.  We assume a ML-SFR scheme is used, the primary band of Cell 0 is also the primary band of Cell 7-12,  and  the secondary band of Cell 1-6.

A UE is placed in Cell 0 and  its position is limited on the straight line between the base station and point A, the intersection of Cell 0,1, and 6.  Denote the  distance between the UE and its base station as

\begin{equation}
\label{ }
r_0=\beta_0 r,
\end{equation}
where $\beta_0$ is a coefficient within $(0,1]$. 

Let us consider the downlink  and assume  $p_n$ be the transmit power density of the base station of Cell $n$, expressed by 

\begin{equation}
\label{ }
p_n=k_n N_0, \quad n=0,1,\cdots,12,
\end{equation}
where $N_0$ is the power density of white noise in the UE receiver.  Assume the bandwidth is $B$, then the power of noise in the UE receiver is 

\begin{equation}
\sigma_z^2=N_0B.
\end{equation}

Denote the distance between the base station of Cell $n$ and the UE as $d_n$  and path loss model as $L(d)$, then the received power of the UE from its serving cell is

\begin{equation}
\sigma_s^2=\frac{p_0 B}{L(d_0)}=\frac{k_0}{L(d_0)}\sigma_z^2,
\end{equation}
the interference power from other cells is 

\begin{equation}
\sigma_I^2=\sum_{n=1}^{12} \frac{p_n B}{L(d_n)}=\sum_{n=1}^{12} \frac{k_n}{L(d_n)}\sigma_z^2.
\end{equation}

Let 
\begin{equation}
k_n=\gamma k_0, \quad n=1,2,\cdots, 6,
\end{equation}

\begin{equation}
k_n=k_0, \quad n=7,8,\cdots, 12,
\end{equation}
meaning the transmit power of all primary bands is $p_0B$, and that of all secondary bands is $\gamma p_0B$, then

\begin{equation}
\sigma_I^2=\left[\gamma \sum_{n=1}^{6}\frac{k_0}{L(d_n)}+ \sum_{n=7}^{12}\frac{k_0}{L(d_n)}\right]\sigma_z^2 .
\end{equation}

Suppose the intra-cell interference is effectively eliminated, as in the OFDM systems,  according to Shannon's law of channel capacity, the maximum spectrum efficiency in a flat fading channel can be expressed as
\begin{eqnarray}
\label{ }
\eta(\gamma,\beta_0)&=&\log_2\left(1+\frac{\sigma_s^2}{\sigma_I^2+\sigma_z^2}\right),
\end{eqnarray}
which is a  function of $\gamma$ and $\beta_0$.

\begin{table}[ht]
  \centering 
  \caption{Calculation parameters }\label{Calparameters}
  \begin{tabular}{l|l}
\hline
  $N_0$(dBm/Hz)& -169\\
  $p_0$(dBm/MHz)& 50/20\\
  $r$(km)& 1\\
  Path loss(dB)&$L(d)=128.1+37.6\log_{10}(d)$ \\
\hline
\end{tabular}
\end{table}

\begin{figure}[!ht]
\begin{center}
\includegraphics[width=3in]{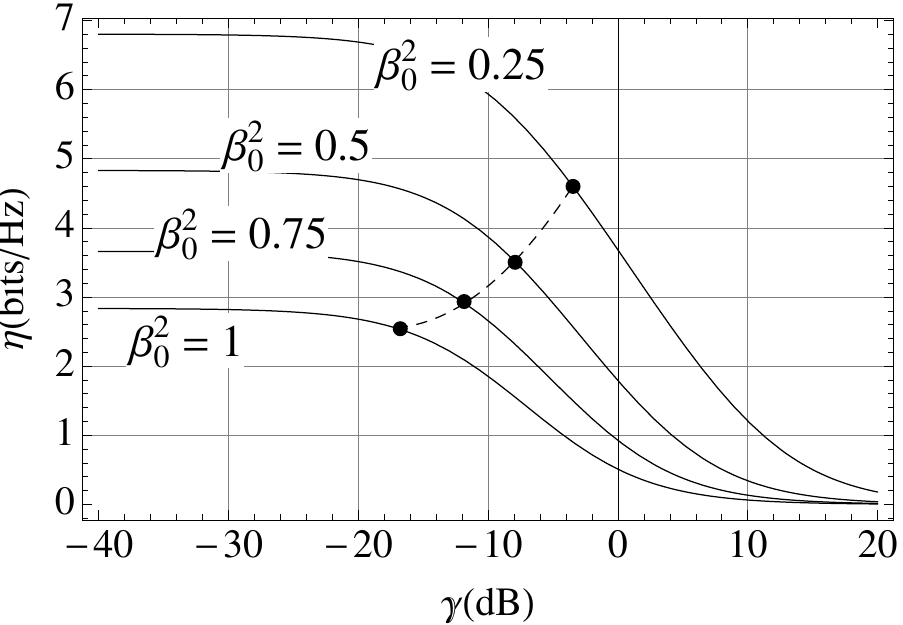}
\caption{Spectrum efficiency  as a function of  $\gamma$ with $\beta_0^2$=0.25,0.5, 0.75 and 1 in a flat fading channel.}
\label{Fig-PTx2}
\end{center}
\end{figure}

We use the parameters in Tab. \ref{Calparameters} and depict $\eta(\gamma,\beta_0)$  as a function of  $\gamma$ with $\beta_0^2$=0.25, 0.5, 0.75 and 1 in Fig. \ref{Fig-PTx2} for a flat fading channel. We can see the curves go down  when $\gamma$ increases due to the increase of inter cell interference.

In engineering application, there may be different options for $\gamma$ parameter.  In our example shown by the  ascending line in Fig. \ref{Fig-PTx2},  the spectrum efficiency for  $\beta_0^2=1$ is selected as about 90\% of the highest value,  and  higher spectrum efficiency is selected for smaller  $\beta_0$ values.  The results of $\gamma$ values are about -17dB, -12.5dB, -8dB, -3dB, based on which a SFR-8 scheme is designed, with the normalized PDL gain of each level  shown in Tab. \ref{SFRGain }. 

\begin{table}[ht]
  \centering 
  \caption{Parameters of  SFR-8}\label{SFRGain }
  \begin{tabular}{c|cccccccc}
\hline
% after \\ : \hline or \cline{col1-col2} \cline{col3-col4} ...
   Level&1&2&3&4&5&6&7&8  \\ \hline
   Gain(dB)& 0 & -2.4& -4.8& -7.3 & -9.7 &-12.1&  -14.6 & -17 \\
\hline
\end{tabular}
\end{table}

\begin{figure}[!ht]
\begin{center}
\includegraphics[width=3in]{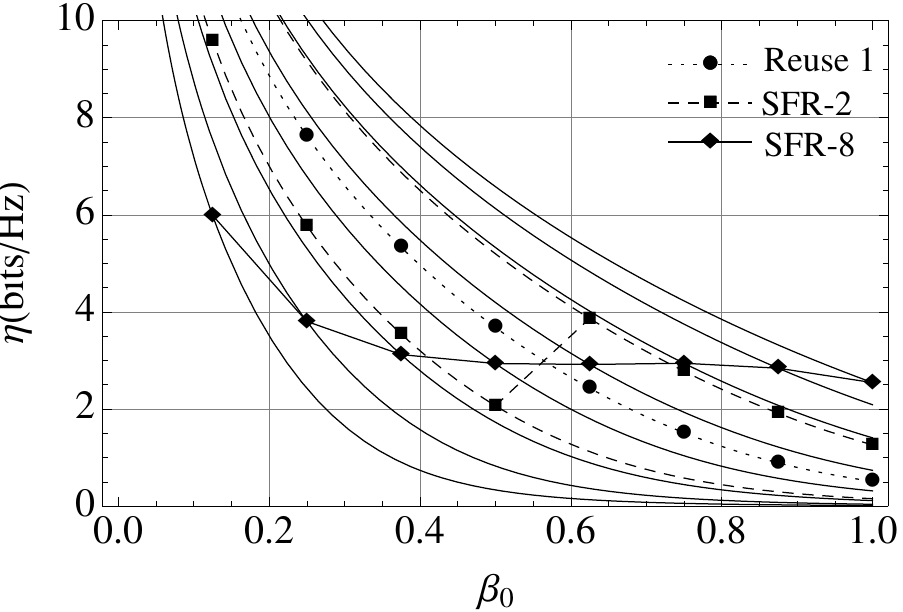}
\caption{Spectrum efficiency as a function of $\beta_0$ of each level of  SFR-8, SFR-2 and reuse 1 for  a flat fading channel.}
\label{Fig-AWGN}
\end{center}
\end{figure}

With such parameters, the spectrum efficiency as a function of $\beta_0$ of each level of  SFR-8, together with SFR-2 and reuse 1 are depicted in Fig. \ref{Fig-AWGN}  for a flat fading channel. The $\gamma$ value for SFR-2 is chosen as -6dB.  

Consider eight positions with $\beta_0=i/8, i=1,2,\cdots,8$, an example resource allocation could be represented by the dots  in Fig. \ref{Fig-AWGN}. For the reuse 1 scheme, high spectrum efficiency can be achieved in cell center, while at  cell edge ($\beta_0=1$), it is only 0.51bps/Hz. SFR-2 improves the cell edge efficiency by 147\% to 1.26bps/Hz,  at the cost of decrease of efficiency in cell centre. SFR-8 realizes a more flat curve and further increases the cell edge efficiency to 2.54bps/Hz, 5 times of that of reuse 1.

\begin{table}[ht]
  \centering 
  \caption{Resource allocation (\%)}\label{ResAlloc}
  \begin{tabular*}{0.49\textwidth}{>{\centering}p{0.95 cm}|>{\centering}p{0.2cm}|>{\centering}p{0.45cm}>{\centering}p{0.45cm}>{\centering}p{0.45cm}>{\centering}p{0.45cm}>{\centering}p{0.45cm}>{\centering}p{0.45cm}>{\centering}p{0.4cm}p{0.45cm}<{\centering}}%{@{\extracolsep{\fill}} c|c|cccccccc}
\hline
% after \\ : \hline or \cline{col1-col2} \cline{col3-col4} ...
   \multicolumn{2}{c|}{$\beta_0$} &1/8 &2/8& 3/8 & 4/8 &5/8 &6/8 &7/8 &1  \\ \hline
   Reuse 1 &1&1.80 &2.71&3.88&5.62&  8.50&13.7&23.2&40.6\\  \hline
   SFR-2  &1&0& 0 & 0 &0 &0&3.45&11.9 &18.0\\
                &2&2.37&3.94 & 6.41&11.0& 20.2 &22.7& 0 &0 \\ \cline{2-10}
                &T&2.37&3.94 & 6.41&11.0& 20.2 &26.2& 11.9 &18.0  \\                 
                \hline
   SFR-8 &1& 0 &  0& 0& 0 &0 &0& 0 &8.33\\    	                	      
	      &2&0 &  0 & 0& 0 &0 &0& 5.50 & 2.83\\     	
   	      &3&0 &  0 & 0& 0 &0 &2.83& 5.50 & 0\\ 
	      &4&0 &  0 & 0& 0 &0 &8.33&0 & 0\\   
	      &5&0 &  0 & 0& 0 &14.0 &2.65& 0 & 0\\ 
	      &6&0 &  0 &0 &14.5 &2.11&0& 0 & 0\\      
   	      &7&0 & 0 & 14.3&2.40 &0 &0& 0 & 0\\
	      &8&4.52 &  11.2 & 0.96& 0 &0 &0& 0 & 0\\ \cline{2-10}
	      &T&4.52 &  11.2 & 15.2& 16.9 &16.1 &13.8& 11.0 & 11.2\\	      
\hline
\end{tabular*}
\end{table}

To achieve a fair comparison, we assume users are distributed on eight circles with $\beta_0=i/8, i=1,2,\cdots,8$, and require the sum data rate on each circle is same.  For every scheme, we perform resource allocation to maximize this sum data rate. The result of resource allocation in percentage is shown in Tab. \ref{ResAlloc}. We can see 40.6\% of resources are allocated to cell edge users for reuse 1 due to the very low spectrum efficiency. With the improvement of cell edge efficiency, this percentage decreases to 18\% for SFR-2 and 11.2\% for SFR-8 and the overall spectrum efficiency is increased. With such resource allocation, the overall spectrum efficiency is 1.654 bps/Hz for reuse 1, while this value is 1.817 bps/Hz for SFR-2 and 2.168 bps/Hz  for SFR-8, increased by   9.85\% and 31\% based on reuse 1,  respectively.

\section{Conclusions}

The proposed multi-level soft frequency reuse scheme is a  generalization and  refinement of SFR-2. ML-SFR and the suggested resource allocation methodology can achieve better interference pattern than SFR-2, further improving the cell-edge and overall data rate.  It can be used in the   current LTE system and would be a  candidate key technology for future 5G systems.

\bibliographystyle{IEEEtran}
\bibliography{IEEEfull,MLSFR}

\end{document}